\documentclass[12pt]{article}
\usepackage[dvips]{graphicx}
\begin{document}

\title{
From ALE-instanton Moduli to Super Yang-Mills Theories via Branes
}
\author{Kei Ito \footnote{e-mail :
 ito@ks.kyy.nitech.ac.jp or keiito@eken.phys.nagoya-u.ac.jp}\\
Department of Electrical and Computer Engineering \\
Nagoya Institue of Technology, \\
Gokiso, Showa-ku, Nagoya 466 Japan}
\maketitle

\begin{abstract}
A large class of equivalence relations between the moduli spaces of
instantons on ALE spaces and the Higgs branches of supersymmetric Yang-Mills
theories, are found by means of a certain kind of duality transformation
between brane configurations in superstring theories. 4d, N=2 and 5d, N=1
supersymmetric gauge theories with product gauge groups turn out to
correspond to the ALE-instanton moduli of type II B and type II A
superstring theories, respectively.

\vspace{30pt}

\end{abstract}

\newpage

In a previous paper[1], N. Maru and the present author constructed brane
configurations in the space-time of superstring theories, which are
equivalent to matrix string theories in the presence of $k$ longitudinal
fivebranes. The low energy limits of these brane confingurations give (1+1)
dimensional $N=(4,4)$ supersymetric $U(n)$ gauge theories with an adjoint
and a singlet hypermultiplet and $k$ matter hypermultiplets in the
fundamental representation of the gauge group.

The latter theory had been studied by Witten[2] and it had been shown that
the Higgs branch of the theory(with zero bare mass and FI terms) can be
interpreted as the moduli space of $n$-instanton solutions of $U(k)$ gauge
theory on ${\bf R}^{4}$, from a point of view of D-flatness of
supersymmetric gauge theories and ADHM construction of instantons.

The moduli space of $n$-instanton solutions of $U(k)$ gauge theory on ${\bf R%
}^{4}$, also admits brane representation. It is a bound state of $n$
Dirichlet(D)-one branes and $k$ D 5 branes, in the type II B superstring
theory, due to Douglas[3].

Since both Higgs branch of supersymmetric gauge theories and the moduli
space of instantons in the gauge theories on ${\bf R}^4$, have their own
brane representations, it is natural to expect that the equivalence relation
between the Higgs branch of (1+1)-dimensional supersymmetric gauge theory
and the moduli space of the instanton solutions of gauge theory, can be
proved more directly at the level of brane configurations.

At first sight, our brane configurations and those of Douglas look quite
different since the former model contain D2 branes and D4 branes, while the
latter contain D1 and D5 branes. There is, however, simple relation between
two types of brane configurations. This is the T-duality transformation in
the $x^6$ direction. (See also ref. [4] , for a discussion of this kind of
duality).

In our model, the world volume of D2 branes, is $(x^0, x^1, x^6)$ and that
of D4 branes, is $(x^0, x^1, x^7, x^8, x^9)$, and $x^6$ direction is
compactified on a circle of radius S. Consider the small radius limit, $S
\to 0$. This limit can be taken by the T-duality transformation of the
branes in the $x^6$ direction. Since the world volume of D2 branes contains $%
x^6$ direction, it loses $x^6$ component upon the T-duality transformatyion
to become $(x^0, x^1)$ which is identified with that of D1 brane. The world
volume of D4 branes, on the contrary, does not contain $x^6$ direction and
so it acquires $x^6$ direction upon thee T-duality transformation to become $%
(x^0, x^1, x^6, x^7, x^8, x^9)$ which is identified with that of D5 brane.

Now in our brane configuration, Higgs branch of the supersymmetric gauge
theory arises from the sector where D2 branes are connected to the D4
branes, which upon T-duality transformation in $x^6$ direction, is mapped on
the sector where D1 branes are bounded to D5 branes, whereby reproducing
Douglas' brane configuration.

A question arises: can one find equivalence relations between the Higgs
branch of supersymmetric theories and the moduli space of instantons,in more
general examples, by considering them in brane configuration and performing
T-duality transformation?

This is the problem which we address in this paper. A natural generalization
of the instanton solution of gauge theory on ${\bf R}^{4}$ (flat space) is,
the instanton solution of gauge theory on a curved four-dimensional
Euclidean space. Among them are the ALE (asymptotically locally Euclidean )
spaces, or Gibbons-Hawking gravitational instanton[5]. So, in this paper, we
try to find supersymmetric gauge theories, the Higgs branch of which are
equivalent to the moduli spaces of instantons of ALE spaces. We will find a
large class of equivalence relations between two sets of theories, by means
of T-duality transformations between the corresponding brane configurations.

The brane configuration which gives the moduli space of $U(k)$ gauge
instanton, with instanton number $n$, on the ALE space with the metric
asymptotic to ${\bf R}^{4}/Z_{m}$ is discussed by Douglas and Moore[6]. In
the case of type II B superstring, it is a bound state of $n$D3 branes with $%
k$D7 branes which are placed at the singularity of ${\bf C}^{2}/Z_{m}$
orbifold Let the world volme of D-branes be;

$nm$ D3 branes: $(x^{0},x^{1},x^{2}.x^{3})$

$k$ D7 branes: $(x^{0},x^{1},x^{2},x^{3},x^{6},x^{7},x^{8},x^{9})$

Then the four-dimensional Euclidean space in question is parametrized by $%
(x^{6},x^{7},x^{8},x^{9})$, so ${\bf C}^{2}$ is parametrized by two complex
numbers; 
\begin{equation}
\left\{ 
\begin{array}{c}
z^{1}=x^{6}+ix^{7} \\ 
z^{2}=x^{8}+ix^{9}
\end{array}
\right. 
\end{equation}

and the $Z_{m}$ action on these numbers is; 
\begin{equation}
\left\{ 
\begin{array}{c}
z^{1}\rightarrow e^{\frac{2\pi i}{m}}z^{1} \\ 
z^{2}\rightarrow e^{-\frac{2\pi i}{m}}z^{2}
\end{array}
\right. 
\end{equation}

The $A_{m-1}$ singularity of ${\bf C}^{2}/Z_{m}$ is at the point $%
(z^{1},z^{2})=(0,0)$, which is $x^{6}=x^{7}=x^{8}=x^{9}=0$. The resolved ALE
space has a metric of Gibbons-Hawking gravitational instanton[5], [6]. 
\begin{equation}
ds^{2}=V^{-1}\left( dt+\vec{A}d\vec{x}\right) ^{2}+V\left( d\vec{x}\right)
^{2}
\end{equation}
\[
V=\sum_{j=1}^{m}\frac{1}{\left| \vec{x}-\vec{x}_{j}\right| }
\]
\[
-\vec{\nabla}V=\vec{\nabla}\times \vec{A}
\]
\[
x^{6}=t,\;\vec{x}=\left( x^{7},x^{8},x^{9}\right) 
\]

For our purpose, the $A_{m-1}$ singularity of ${\bf C}^{2}/Z_{m}$ can be
conveniently represented by $m$ copies of NS five branes with the world
volume;

$m$ NS5 branes: $(x^{0},x^{1},x^{2},x^{3},x^{4},x^{5})$

since these branes have definite $(x^6, x^7, x^8, x^9)$ value corresponding
to their position coordinates.

Now under the T-duality transformation in $x^6$ direction, branes are
transformed as follows;

$nm$ D3 branes $(x^{0},x^{1},x^{2},x^{3})$

${\to }$ $n$ D4 branes $(x^{0},x^{1},x^{2},x^{3},x^{6})$

$k$ D7 branes $(x^{0},x^{1},x^{2},x^{3}.x^{6}.x^{7}.x^{8}.x^{9})$

${\to }$ $mk$ D6 branes $(x^{0},x^{1},x^{2},x^{3},x^{7},x^{8},x^{9})$

since D3 branes acquire $x^{6}$ component, and D7 branes lose $x^{6}$
component in the world volume. On the other hand, m NS5 branes with the
world volume $(x^{0},x^{1},x^{2},x^{3},x^{4},x^{5})$ are not changed under
the duality transformation. The brane configuration after the duality
transformation is depicted in fig. 1. 
\begin{figure}[htbp]
\includegraphics[height=7cm]{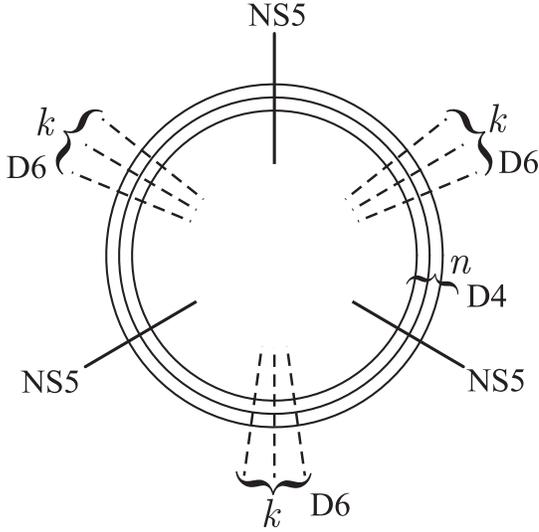}
\caption{Type II B theory is transformed into type II A theory and gives
this brane configuration of $n$ D4, $mk$ D6 and $m$ NS5 branes. This figure
is drawn for $n=m=k=3$. The effective field theory is a 4-dimensional
supersymmetric Yang-Mills theory.}
\label{fig1}
\end{figure}
From this configuration, one can read
off the effective supersymmetric gauge theory, to which it reduces at low
energies. The world volumes of D4 branes are $%
(x^{0},x^{1},x^{2},x^{3},x^{6}) $, but they have only finite length in $x^{6}
$ direction, so macroscopically, the D4-brane effective field theory is on
the four-dimensional space-time $(x^{0},x^{1},x^{2},x^{3})$. In the course
of T-duality transformation, the original type II B string is transformed
into type II A string, which has N=2 supersymmetry in ten dimensions. This
is equivalent to N=8 supersymmetry in four dimensions. In the presence of
NS5 branes, D4 branes and D6 branes, one quarter of the original
supersymmetry is preserved, so it gives N=2 supersymmetry in four
dimensions. The presence of $n$ D4 branes gives $U(n)$ gauge group, but they
are suspended between $m$ NS5 branes located at equal spacing $2\pi /m$ in
the circle of radius S in $x^{6}$ direction. Therefore, the gauge group
becomes a product of $U(n)$ groups, which is $U(n)^{m}$. $mk$ D4 branes give
matter hypermutiplets, first $k$ of which are in the $(n,\underbrace{%
1,1,\cdots ,1}_{m-1})$ representation of $U(n)^{m}$ next $k$ of which are in
the $(1,n,\underbrace{1,\cdots ,1}_{m-2})$, and so on. The last $k$ hyper
multiplets are in the $(\underbrace{1,\cdots ,1}_{m-1}n)$ representation of $%
U(n)^{m}$. There arise also $m$ hypermultiplets in the bi-fundamental
representations of $U(n)^{m}$, which are $(n,\bar{n},\underbrace{1,\cdots ,1}%
_{m-2}),$ $(1,n,\bar{n},\underbrace{1,\cdots ,1}_{m-3}),$ $\,\cdots $ $\,,(%
\bar{n},\underbrace{1,\cdots ,1}_{m-2},n).$ Now we proceed to the instanton
moduli on ALE in the type II A superstring. In type II A superstring, the
brane configuration of Douglas and Moore is, a bound state of $n$ D4branes
with $k$ D 8 branes, as opposed to $n$ D3 branes with $k$ D7 branes in the
type II B case. Under the T-duality transformation in $x^{6}$ direction,
branes are transformed as follows;

$nm$ D4 branes $(x^{0},x^{1},x^{2},x^{3},x^{4})$ ${\to }n$ D5 branes $%
(x^{0},x^{1},x^{2},x^{3},x^{4},x^{6})$

$k$D8 branes $(x^{0},x^{1},x^{2},x^{3},x^{4},x^{6},x^{7},x^{8},x^{9})$ ${\to 
}$

$mk$ D7 branes $(x^{0},x^{1},x^{2},x^{3},x^{4},x^{7},x^{8},x^{9})$

On the other hand$\ m$ NS 5branes with the world volume $%
(x^{0},x^{1},x^{2},x^{3},x^{4},x^{5})$ are not changed under the duality
transformation. The brane configuration after the duality transformation is
depicted in fig. 2.
\begin{figure}[htbp]
\includegraphics[height=7cm]{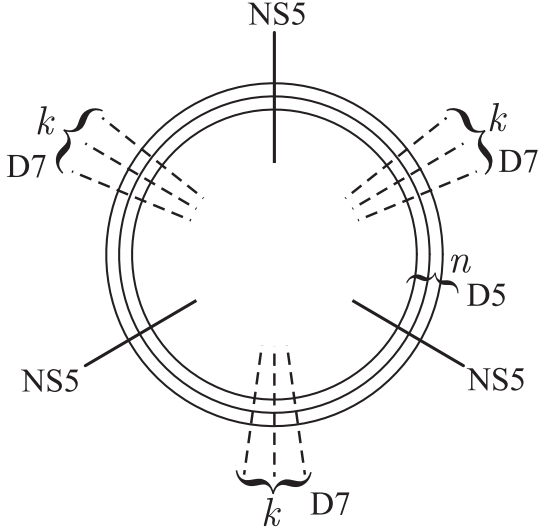}
\caption{Type IIA theory is transformed into type II B theory and gives
this brane configuration. The effective field theory in this case is 5
dimensional.}
\label{fig2}
\end{figure}
 The crucial difference of this type IIA case (IIB after
the duality transformation) from the above type II B case (IIA after the
duality transformation) is that the dimension of the space-time of the
low-energy effective field theory is 5, instead of 4, since there are D5
branes with world volume $(x^{0},x^{1},x^{2},x^{3},x^{4},x^{6})$, which
looks macroscopically like the five-dimensional space-time $%
(x^{0},x^{1},x^{2},x^{3},x^{4})$. Therefore, we have five-dimensional N=1
supersymmetric gauge theory as an effective theory. The gauge group and
hypermultiplets and their representations are the same as in the case of
type II B(II A after the duality transformation).

In conclusion, we have found a large class of equivalence relations between
the moduli spaces of instantons on ALE spaces and the Higgs branches of
supersymmetric Yang-Mills theories, by meaans of T-duality transformation in 
$x^6$ direction between brane configurations in type II A and type II B
superstring theories.

The equivalence relations we have found are summarized as follows;

The moduli spaces of $U(k)$ gauge instanton with instanton number $n$ on the
ALE space with the metric asymptotic to ${\bf R}^{4}/Z_{m}$ in type IIB
(type II A) superstring is, isomorphic to the Higgs branch of 4-dimensional
N=2 ( 5-dimensional N=1 ) supersymmetric $U(n)^{m}$ Yang-Mills theory with $%
mk$ hypermultiplets in $(n,1,\cdots ,1)$, $(1,n,1,\cdots ,1)$, ...... , $%
(1,1,\cdots ,n)$ representations of $U(n)^{m}$ and $m$ hypermultiplets in
the bi-fundamental representations of $U(n)^{m}$, $(n,\bar{n},1,\cdots ,1)$, 
$(1,n,\bar{n},1,\cdots ,1)$, ..... , $(\bar{n},1,\cdots ,1,n)$.

Acknowledgements

The author would like to thank S. Kitakado, H. Ikemori and N. Maru for useful
discussions.

\newpage
REFERENCES

\noindent
[1] Kei Ito and Nobuhito Maru, "Matrix String Theory from Brane
configuration" hep-th/9710029.

\noindent
[2] E. Witten, "On the Conformal Field Theory of the Higgs Branch" ,
hep-th/9707093.

\noindent
[3] M. R. Douglas, "Branes within Branes", hep-th/9512077.

\noindent
[4] J. H. Brodie, "Two Dimensional Mirror Symmetry from M-theory",
hep-th/9709228.

\noindent
[5] G. W. Gibbons and S. W. Hawking, Commun. Math. Phys. 66(1979)291.

\noindent
[6] M. R. Douglas and G. Moore, "D-branes, Quivers, and ALE Instantons"
hep-th/9603167

\end{document}